# Mechanical properties of $B_{12}$-based orthorhombic metal carboborides. A first principle study.

Oleksiy Bystrenko[a,c], Tianxing Sun[a,b], Zhaohua Luo[a,b], Jingxian Zhang[a,b], Yusen Duan[a,b], Kaiqing Zhang[a,b], Hu Ruan[a,b], Wenyu Tang[a,b], Bohdan Ilkiv[c], Svitlana Petrovska[c], Tetiana Bystrenko[c]

[a] Key Laboratory of Advanced Structural Ceramics and Ceramics Matrix Composites, Shanghai Institute of Ceramics, Chinese Academy of Sciences, Shanghai 200050, China

[b] Center of Materials Science and Optoelectronics Engineering, University of Chinese Academy of Sciences, Beijing 100049, China

[c] Frantsevich Institute for Problems of Materials Science, National Academy of Science of Ukraine, Kyiv, Ukraine

Abstract:
Structural and mechanical properties of $B_{12}$-based orthorhombic metal carboborides are studied on the basis of first principle DFT approach. The simulations predict the existence of a new family of phases of the composition $MeC_2B_{12}$ (Me=Mg, Ca, Sr, Sc, Y) with similar structure and space symmetry Imma. It has been found that the predicted phases are thermally (dynamically) stable and have considerably better mechanical properties as compared to the reference compound $AlMgB_{14}$. The respective calculated isotropic elastic moduli and Vickers hardness are significantly higher (G~ 230-250, E~530-550, and Hv~35-55 GPa). These conclusions were confirmed by direct calculations of shear strength for the above phases, which demonstrated the increase of 30-50% in different directions. The enhanced mechanical characteristics of the $MgC_2B_{12}$-based phases make them promising for creating novel superhard materials.

Keywords:     hard materials, boron icosahedra, mechanical properties, orthorhombic metal borides, DFT

## I. Introduction

Orthorhombic metal borides and related compounds based on icosahedral $B_{12}$ structural units attracted significant attention of researchers during last decades. A distinguishing feature of compounds of this family is high hardness, which makes them promising for creating novel superhard materials. The basic well examined reference phase in this context is $AlMgB_{14}$ (space group Imma [74]) with the hardness 28-35 GPa [1, 2]. It has low density (2.7 g/cm3), is inexpensive to produce, and has found already the use in industrial applications. Its structure is represented by boron layers formed by $B_{12}$ structural units with Al and Mg atoms embedded on metal sites (Fig. 1, on the left) [3].

To find related compositions with better mechanical properties, a number of simulations has been done (see, for instance, [4-9]). The results of these latter indicate that a simple replacement of Al and Mg atoms on metal sites by different elements can not essentially improve mechanical characteristics of the respective compounds suggesting that the limiting factor is the structure of boron framework itself rather than specific elements on metal sites. At the same time, significant increase of elastic moduli has been predicted by first principle simulations for a closely related compound, $MgC_2B_{12}$ (Imma [74]), which was synthesized earlier [10]. Its crystal structure can be viewed as a modification of $AlMgB_{14}$ lattice obtained by removing the aluminum from metal sites and replacing the inter-icosahedral boron atoms by carbon (Fig. 1). For this reason, it was of interest to examine theoretically the mechanical properties and stability of a family of related structures with Mg replaced by different metals.

A theoretical study of the related phases with the compositions $MeC_2B_{12}$, Me= Mg, Ca, Sr was performed in the previous work of the authors [11]. It has been found that the theoretically predicted Ca- and Sr- based phases with the structure similar to that of the compound $MgC_2B_{12}$ (with the same Wickoff positions of ions and space symmetry Imma) are mechanically and dynamically stable and have considerably better mechanical properties, in particular, significantly higher elastic moduli. The Vickers hardness was estimated to be in the range 45-55 GPa. However, the empirical models for hardness employed in this work have the drawback that they evaluate hardness on the basis of elastic properties of materials, which are defined for small strains. In actual fact, however, hardness is associated with much larger plastic deformations. In view of this, it is important to verify the predictions for hardness of the phases under consideration by means of alternative approaches. In particular, the examination of shear strength of compounds, which is directly related to plastic processes, can be viewed as additional reliable way to make conclusions about hardness.

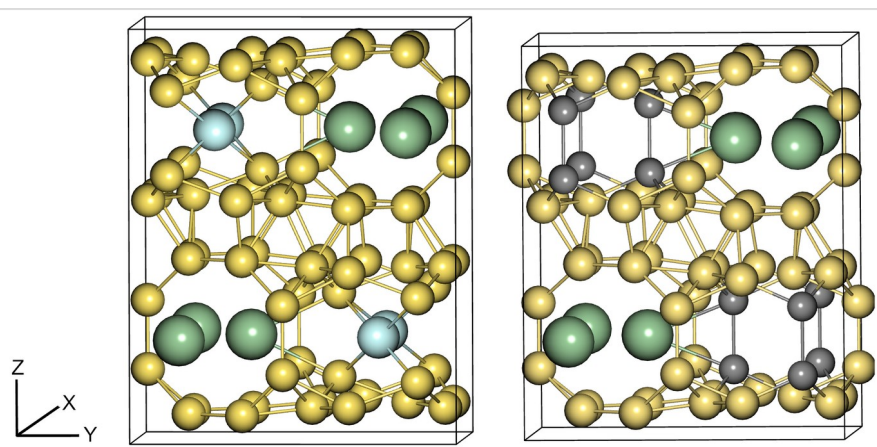

Figure 1. Structure of $AlMgB_{14}$ with 64 atoms in unit cell (left), and $MgC_2B_{12}$ with 60 atoms in unit cell (right). Aluminum, magnesium, boron, and inter-icosahedral carbon atoms are given in silver, green, yellow and gray, respectively. The structure of $MgC_2B_{12}$ can be obtained from $AlMgB_{14}$ by replacing the inter-icosahedral boron atoms by carbon and removing the aluminum atoms.

Thus, the primordial purpose of this work is to verify the results of Ref. [11] concerning the hardness of predicted $MgC_2B_{12}$ related phases by direct calculations of shear strain-stress behavior and evaluation of shear strength in different directions. In addition, we included into consideration two more related phases with similar structure by using the elements of III group, Sc and Y, as metal atoms. The study was based on first principle computer simulations aimed predominantly at structural and mechanical properties and stability of the above mentioned theoretically predicted phases.

## II. Computational methods

First principle simulations were carried out on the basis of density functional theory (DFT) implemented in Quantum Espresso software [12]. For calculations, the pseudopotentials from SSSP 1.1.2 library [13] employing the generalized gradient approximation (GGA) with the exchange-correlation energy in Perdew–Burke-Ernzerhof form [14] were used. Sampling of the Brillouine zone was done on the uniform Monkhorst-Pack grids [15] of sizes 5x5x5, 5x3x3 and 4x2x2 with no shift. The cutoff energy for the basis wave functions and the charge was set 55 Ry and 440 Ry, respectively, in all cases, except for the Sc-based phase, for which the respective numbers were 90 Ry and 720 Ry; the total energy convergence threshold was set $10^{-6}$ Ry. Simulations of the relaxation of the structures were carried out by using BFGS algorithm [16] with the threshold $10^{-4}$ Ry/Bohr for ions. Identification of crystal system and space symmetry of the phases obtained in simulations was done by means of FINDSYM software [17].

Elastic properties of the phases under consideration were evaluated within the strain-stress approach implemented in THERMO_PW software [18]. The latter calculates stress responses for a series of strains, which are used then to evaluate the elastic constants for the selected crystal structure. The elastic moduli of respective policrystalline materials can be found after that by using Voigt, Reuss or Hill averaging approaches [19].

Theoretical prediction of hardness represents a challenging task, since this is a macroscopically defined quantity. This means that it depends on a large number of conditions (like the number of defects, porosity, etc.) which makes it impossible to unambiguously deduce hardness from first principle simulations. In this work we obtained the estimates for hardness from simple empirical approaches based on the experimentally observed correlation between hardness and elastic moduli of materials proposed in Ref. [20].

## III. Results of simulations

In simulations, we considered the basic reference compound $AlMgB_{14}$ and phases with the composition $MeC_2B_{12}$, (Me=Mg, Sr, Ca, Sc, Y) with the structure similar to that of the phase $MgC_2B_{12}$ observed experimentally [10]. Initial structural data needed for simulations was taken from our previous works [8, 11], and from the materials database [21].

At first stage, the preliminary configurations for the phases to consider were prepared. The $MgC_2B_{12}$ related phases were constructed by replacing Mg by respective metal atoms in the reference structure. Then, in order to accurately determine the equilibrium configurations associated with energy minima, the relaxation procedure was applied with taking into account all ionic coordinates and cell degrees of freedom including lattice parameters and angles. After that, these phases were examined in simulations for structure, elastic properties and stability.

The structural, mechanical and some other basic properties of the Sc- and Y-based phases obtained in these simulations are given in Tables 1-2. The theoretical results for the Mg,- Ca-, and Sr- based phases and the basic reference compound $AlMgB_{14}$ obtained in the previous works of the authors [8, 11] are given for comparison, along with the experimental data of other relevant studies.

As follows from the results presented, the structures calculated for Sc- and Y-based phases are very similar to those of the $MgC_2B_{12}$ and the early predicted for related Ca-, and Sr- based compounds (Table 1). They all belong to the same orthorhombic crystal system, have the same space symmetry Imma and Wickoff positions for ions [11].

Table 1. Structural and other basic properties of $B_{12}$-based metal carboborides $ScC_2B_{12}$ and $YC_2B_{12}$ obtained in this work and other $MgC_2B_{12}$-based compounds from relevant studies. For comparison, the results for the reference system $AlMgB_{14}$ are given. Theoretical results of the previous works of the authors [8, 11] are marked with *; experimentally observed phases are marked with +. The theoretical and experimental data of other relevant studies are given in round and curly brackets, respectively.

| Composition, $N_{cell}$ | Crystal system, Space group | Lattice parameters, A | | | Cohesive energy per atom (eV) | Formation energy per atom (eV) | Density (g/cm$^3$) |
|---|---|---|---|---|---|---|---|
| | | a | b | c | | | |
| $ScC_2B_{12}$ 60 | Orthorhombic Imma [74] | 5.59 | 7.93 | 9.90 | | -0.244 | 3.01 |
| $YC_2B_{12}$ 60 | Orthorhombic Imma [74] | 5.61 | 8.05 | 10.03 | | -0.232 | 3.56 |
| *$CaC_2B_{12}$ 60 | Orthorhombic Imma [74] | 5.61 | 8.04 | 10.02 | -6.834 | -0.250 | 2.85 |
| *$SrC_2B_{12}$ 60 | Orthorhombic Imma [74] | 5.63 | 8.15 | 10.20 | -6.744 | -0.180 | 3.43 |
| *+$MgC_2B_{12}$ 60 | Orthorhombic Imma [74] | 5.61 {5.613}[c] | 7.93 {7.933}[c] | 9.81 {9.828}[c] | -6.725 (-6.879)[a] | -0.168 (-0.139)[b] | 2.71 |
| *+$AlMgB_{14}$ 64 | Orthorhombic Imma [74] | 5.907 (5.900)[b] | 8.110 (8.114)[b] | 10.346 (10.345)[b] | -6.151 (-6.296)[a] | -0.089 (-0.092)[b] | 2.72 |

[a]Ref.[4]; [b]Ref.[5]; [c]Ref.[10]

Table 2. Mechanical properties of metal carboborides $ScC_2B_{12}$ and $YC_2B_{12}$ obtained in this work and other $MgC_2B_{12}$-based compounds from relevant studies. For comparison, the results for the reference system $AlMgB_{14}$ are given. Theoretical results of the previous work of the authors [8, 11] are marked with *; experimentally observed phases are marked with +. The theoretical and experimental data of other relevant studies are given in round and curly brackets, respectively. The numbers given for elastic moduli are Voigt-Reuss-Hill averages; theoretical Vickers hardness was evaluated from elastic moduli on the basis of the models given in Ref. [20] according to the Eqs. (1) and (2) (in brackets).

| Compo-sition | Elastic moduli (GPa) | | | Pugh's ratio k | Poisson's ratio ν | Debye temperature (K) | Vickers hardness (GPa) |
|---|---|---|---|---|---|---|---|
| | Bulk modulus B | Shear modulus G | Young's modulus E | | | | |
| $ScC_2B_{12}$ | 240.2 | 235.9 | 533.2 | 0.982 | 0.13 | 1486 | 35.6 [44.9] |
| $YC_2B_{12}$ | 237.4 | 248.1 | 551.9 | 1.045 | 0.11 | 1385 | 37.5 [50.1] |
| *$CaC_2B_{12}$ | 224.6 | 247.2 | 542.6 | 1.101 | 0.10 | 1544 | 37.3 [53.0] |
| *$SrC_2B_{12}$ | 214.7 | 248.7 | 538.3 | 1.158 | 0.08 | 1394 | 37.6 [56.9] |
| *+$MgC_2B_{12}$ | 231.2 (228.2)[a] | 218.6 (217)[a] | 498.6 (494.4)[a] | 0.946 | 0.14 (0.139)[a] | 1507 | 33.0 [40.8] |
| *+$AlMgB_{14}$ | 201.2 (199.6)[a] | 193.7 (191.5)[a] | 440.0 (435.3)[a] | 0.963 | 0.14 (0.136)[a] | 1391 | 29.2 [38.8] {32-35}[b] |

[a]Ref.[5]; [b]Ref.[1].

The calculated isotropic elastic moduli and related quantities for the phases under examination are given in Table 2. The most remarkable finding is a significant increase of predicted elastic moduli and hardness of the $MgC_2B_{12}$-related phases as compared to the reference phase $AlMgB_{14}$. At the same time, as can be seen from the Table 2, there is a considerable difference between the theoretical estimates for Vickers hardness made on the basis of two empirical models proposed in Ref. [20],

$$H_V = 0.151G, \quad (1)$$

and

$$H_V = 2(k^2 G)^{0.585} - 3, \quad (2)$$

where k=G/B is the Pugh's ratio. This can be understood from Fig. 2, where the data for a number of hard compounds is given on the shear modulus vs. Vickers hardness plane, along with the relation (1). As is clearly visible from the figure, the compounds with hardness above ~ 50 GPa mostly have the larger Pugh's ratios (greater than 1), whereas the Eg. (1) gives for them typically underestimated numbers. The compounds with lower Pugh's ratio have mostly lower hardness numbers, while the predictions from Eq. (1) are overestimated. Thus, a little more sophisticated model (2) is designed to better reproduce the correlation between hardness and elastic properties by taking into account the Pugh's ratio.

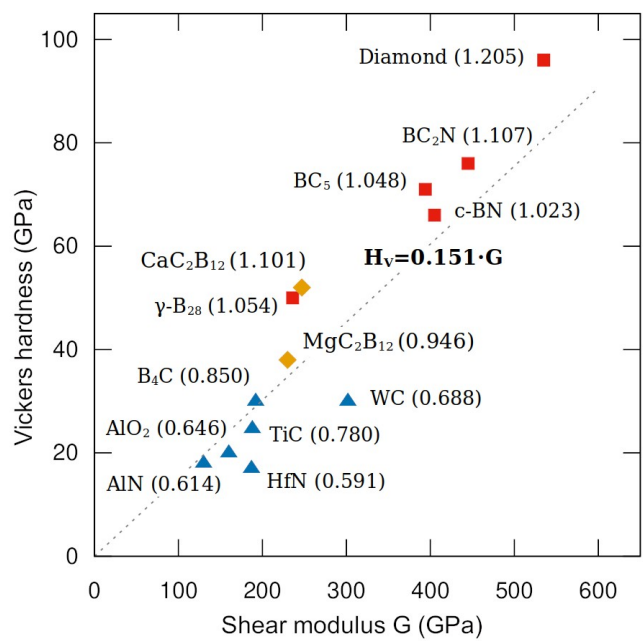

Figure 2. Illustration to the empiric relation for hardness $H_V=0.151G$ (dashed line). The numbers in round brackets are the Pugh's modulus ratios. The data is taken from Ref.[20] or is the results of present simulations.

As mentioned before, it is important to verify the numbers for the hardness obtained from simple empirical models on the basis of alternative approaches. Hardness is a quantity associated with plastic processes and, therefore, is related to shear strength of materials. Thus, this latter can be useful in examining hardness properties. In order to determine the shear strength of the phases under consideration, we calculated the shear strain-stress dependencies in a straightforward manner by using first principle DFT approach. This made the examination of the elastic failure of the respective phases and the evaluation of the respective shear strength possible. The simulations were performed for selected crystal directions for the shear strains up to 0.25-0.3 in the following way. For a given shear system, a series of uniform shear deformations has been successively applied to the unit cell for the selected phase. At each step, the ionic subsystem and lattice parameters were allowed to relax in such a way as to keep the selected shear system unchanged. After that, the resulting shear strain and stress were recalculated.

As has been shown in Refs [8, 9], the hardness of orthorhombic $AlMgB_{14}$-related compounds is limited by the strength of bonds connecting boron layers and the weakest shear systems are associated with relative sliding of boron layers. With the coordinate system defined as in Fig.1, these weakest shear systems can be identified as (001)[100] and (001)[010], while the strongest shear system is expected to be (010)[100]. The results of simulations of shear strain-stress dependencies for these three directions are given in Figs. 3-5, and the information about the corresponding shear strengths is summarized in Table 3.

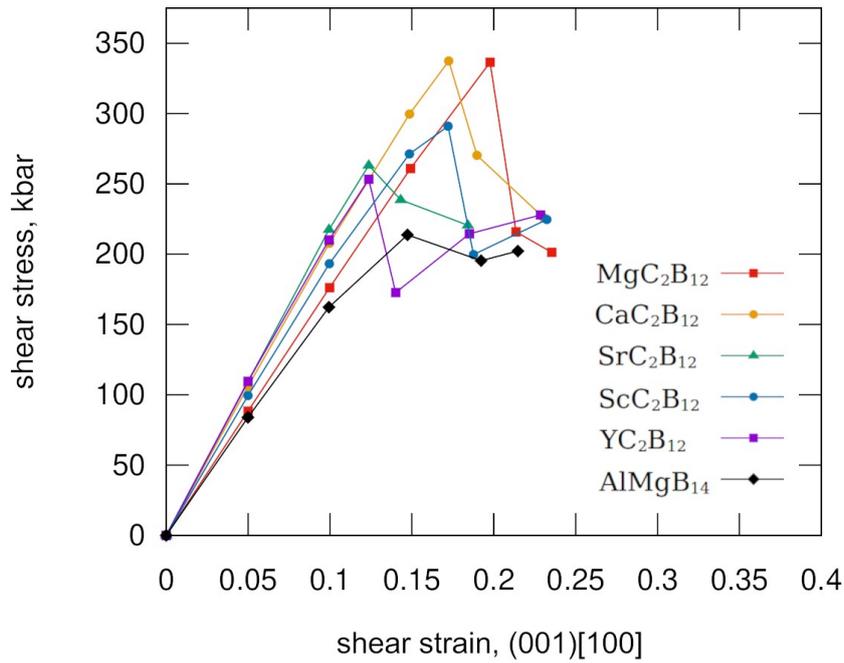

Figure 3. Calculated shear stress-strain relations for the first weak shear system, (001)[100]. The shear strain is defined as $\varepsilon=\Delta c_x/c_z$, where c is the respective lattice dimension.

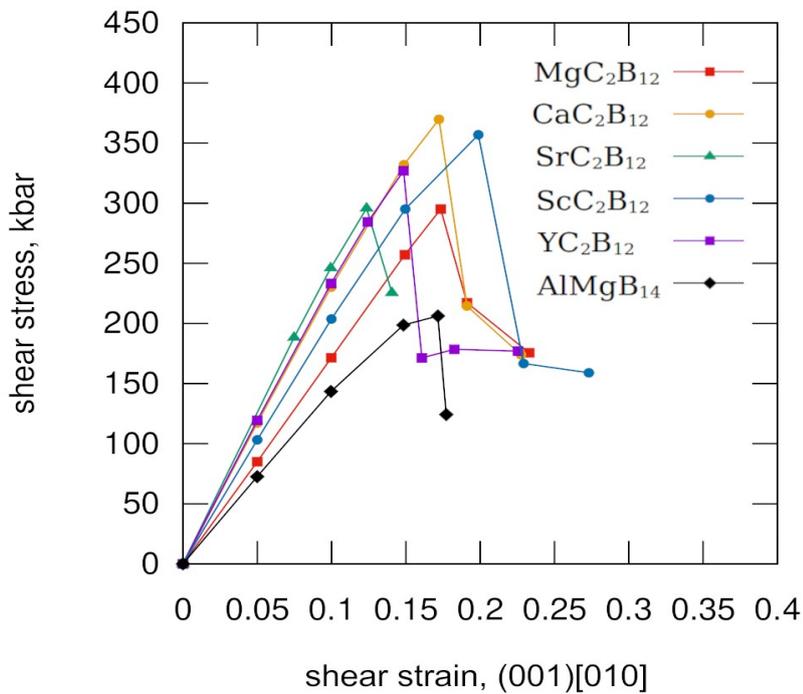

Figure 4. Calculated shear stress-strain relations for the second weak shear system, (001)[010]. The shear strain is defined as $\varepsilon=\Delta c_y/c_z$, where c is the respective lattice dimension.

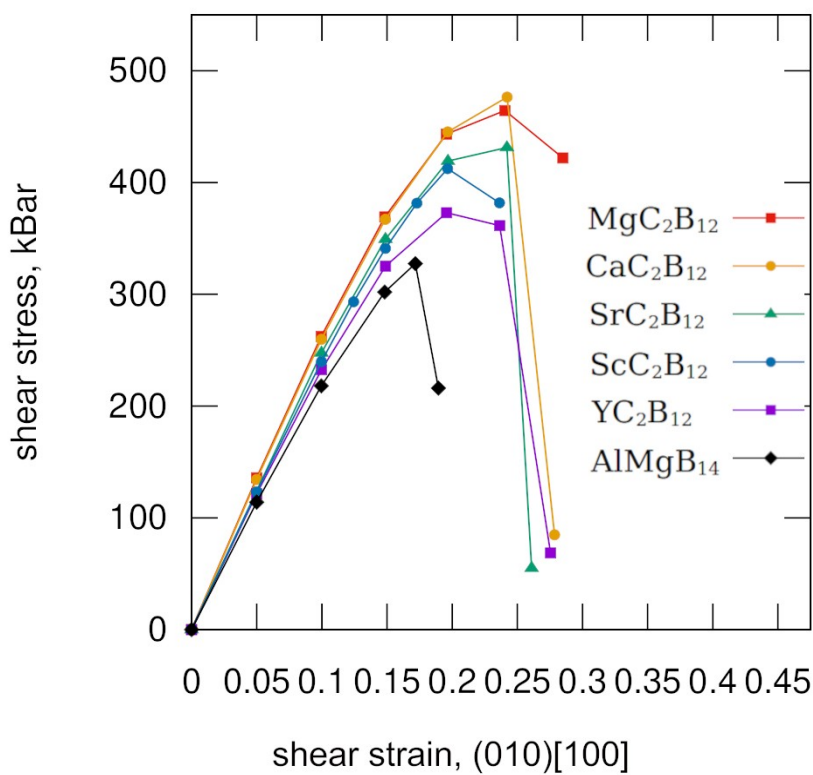

Figure 5. Calculated shear stress-strain relations for the strongest shear system, (010)[100]. The shear strain is defined as $\varepsilon=\Delta b_x/b_y$, where b is the respective lattice dimension.

Table 3. Shear strength of orthorhombic metal carborides in comparison with $AlMgB_{14}$ for the weakest (001)[100] and (001)[010], and the strongest (010)[100] shear systems calculated in this work.

| Shear system | Shear strength, GPa | | | | | |
|---|---|---|---|---|---|---|
| | $MgC_2B_{12}$ | $CaC_2B_{12}$ | $SrC_2B_{12}$ | $ScC_2B_{12}$ | $YC_2B_{12}$ | $AlMgB_{14}$ |
| (001)[100] | 33.6 | 33.7 | 26.3 | 29.3 | 25.3 | 21.4 (~20)[a] |
| (001)[010] | 29.5 | 37.0 | 29.5 | 35.7 | 32.7 | 20.6 |
| (010)[100] | 46.4 | 47.6 | 43.1 | 41.3 | 37.3 | 32.7 (~32)[a] |

[a] The numbers in round brackets are estimated from the graphical data of Ref. [5].

The data given in the Figures 3-5 and Table 3 clearly supports the conclusions about the high hardness of the phases under consideration. All the $MgC_2B_{12}$-related phases considerably exceed the reference compound $AlMgB_{14}$ in shear strength. The most remarkable increase of ~50% in all three directions is observed for Mg- and Ca- based phases.

The shears (001)[100] and (001)[010] associated with relative sliding of boron layers turn out to be indeed much weaker than (010)[100] shear. As mentioned above, this can be the result of relatively weak bonding between boron layers. To examine this point in more detail, we calculated the changes in electron density during the elastic failure in the plane (100) containing the interlayer bonds connecting the inter-icosahedral boron or carbon atoms for AlMgB$_{14}$ or MgC$_2$B$_{12}$.

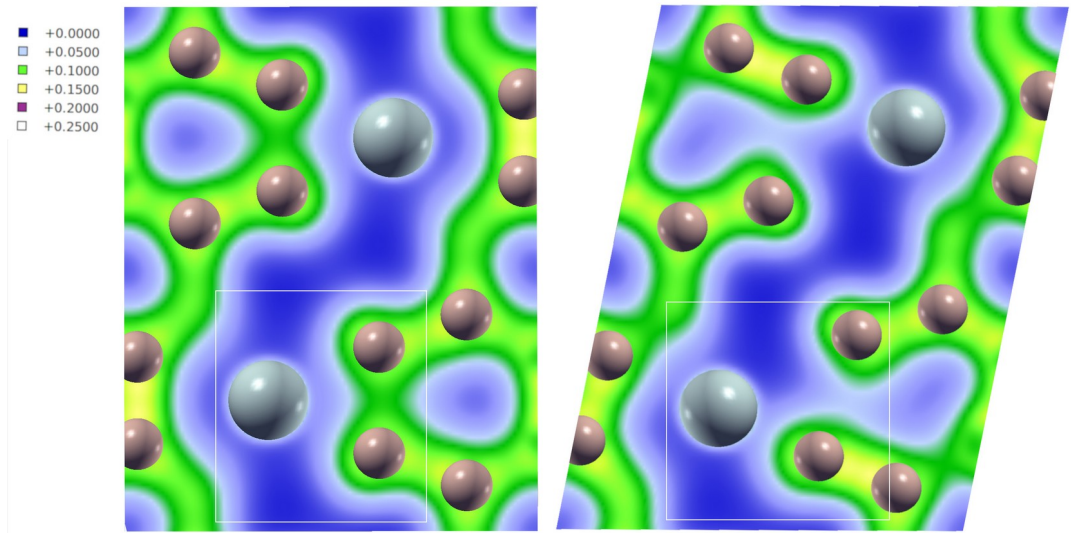

Figure 6. Changes in electron charge density in AlMgB$_{14}$ associated with the elastic failure occuring within the strain range ε=0.175-0.2 for the weak shear system (001)[010] describing the relative sliding of boron layers. The breaking of interlayer B-B bonds can be observed.

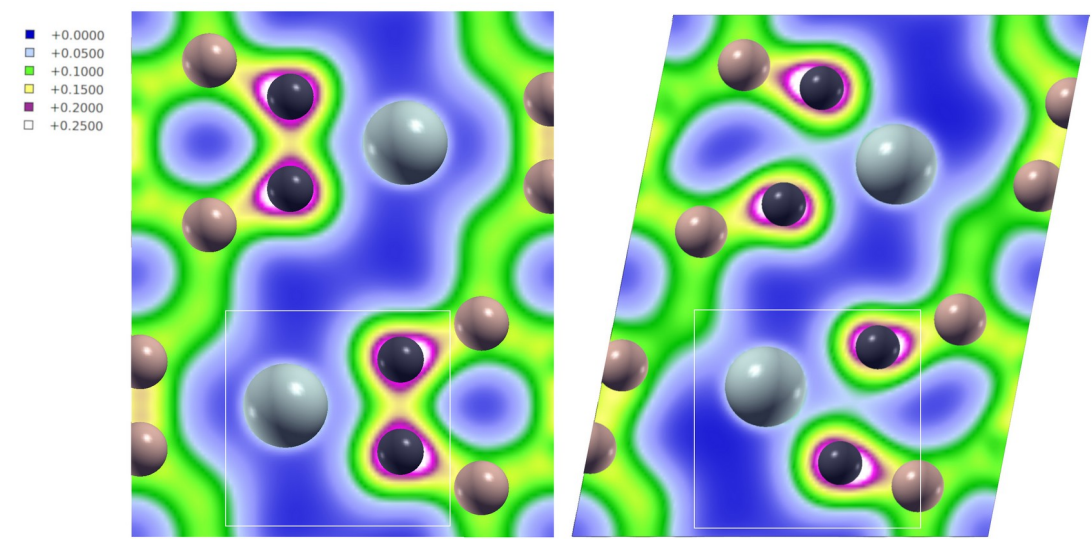

Figure 7. Same as in Fig.6 but for the phase MgC$_2$B$_{12}$. The breaking of interlayer C-C bonds can be observed.

The results (Figs. 6, 7) indicate that elastic failure occuring within the strain range ε=0.175-0.2 for the weak shear system (001)[010] associated with layer sliding is indeed accompanied by the breaking of the respective B-B or C-C bonds. At this, the greater strength of C-C bonds is reflected by greater electron density. Thus, the results given in Figs. 6, 7 provide the illustration to the conclusions of Refs. [8, 9] concerning the impact of the interlayer bonds on the hardness of orthorhombic metal borides.

Comparison of calculated electronic density of states (DOS) for the phases $ScC_2B_{12}$, $YC_2B_{12}$ and $AlMgB_{14}$ is given in Fig. 8. As can be seen from the figure, the behavior of DOS for the phases containing the metal atoms of III group is similar to that of $AlMgB_{14}$. However, it is to notice that the Mg- and Al-related partial DOS within the valence band for $AlMgB_{14}$ is close to zero, which is consistent with purely ionic nature of bonding of metal atoms in this compound [4]. At the same time, the Sc- and Y- based phases under consideration demonstrate marked contributions to partial DOS of metal atoms in valence bands reflecting the formation of covalent bonds, which may be associated with the change in mechanical properties.

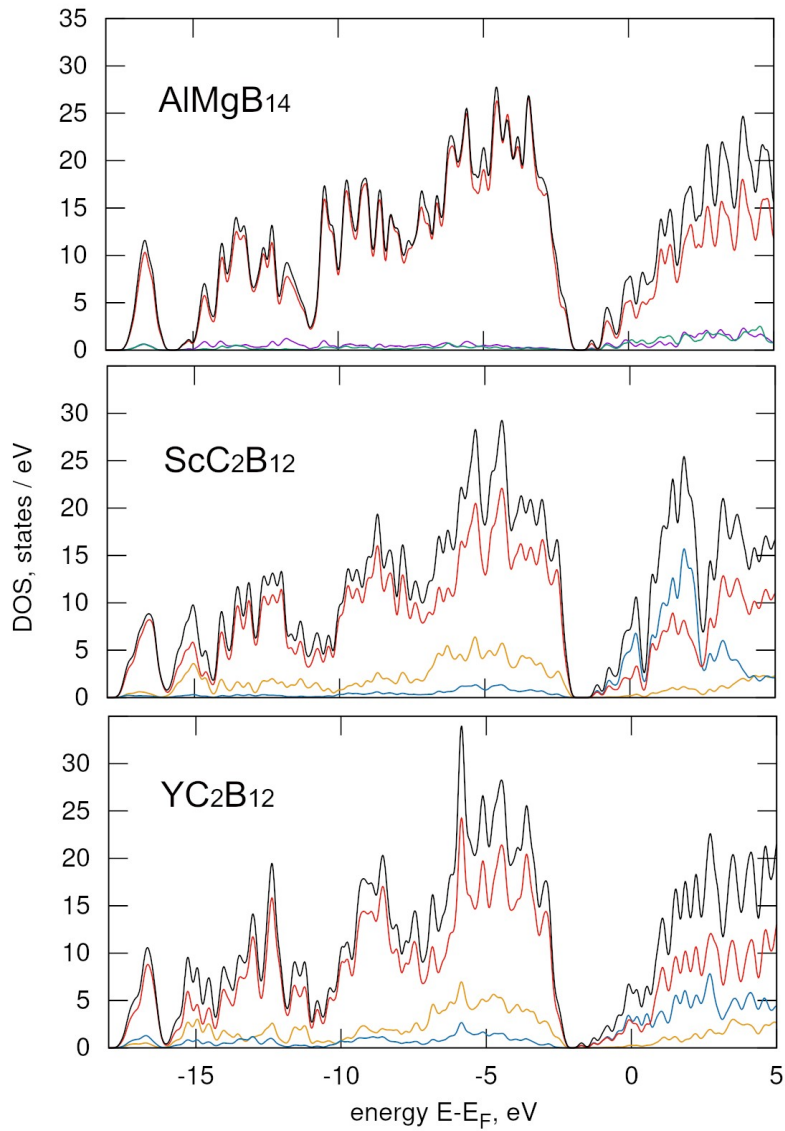

Figure 8. Electronic density of states for $MgC_2B_{12}$-related compounds with the metal atoms of III group in comparison with $AlMgB_{14}$ obtained in simulations. Total DOS is given in black; the curves in red, yellow, and green/violet/blue relate to partial DOS for B, C and metal element, respectively.

There is one more issue concerning the theoretically predicted phases. The matter is that the conventional optimization procedure at zero temperature employed to obtain the equilibrium crystal structures does not in general guarantee the dynamical stability of these latter. Dynamical stability implies that every single degree of freedom (mode) of a multiparticle system is stable with respect to excitation. At non-zero temperature, due to thermalization, every mode acquires some amount of kinetic energy, and the presence of unstable modes results in breakdown of the structure.

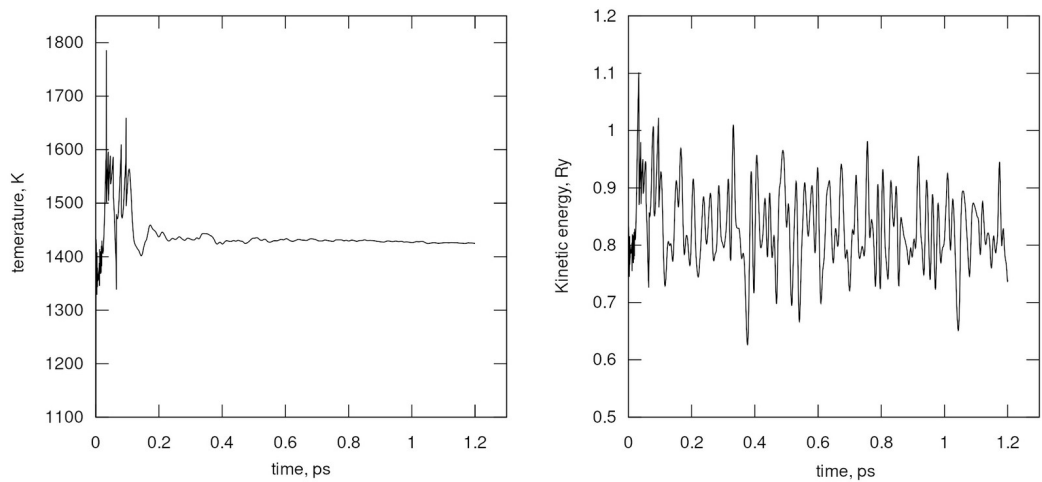

Figure 9. Behavior of temperature and kinetic energy of ion subsystem in the process of thermalization, obtained in Born-Oppenheimer molecular dynamics simulations for $ScC_2B_{12}$ compound. The initial temperature was set T~1480 K; the steady state temperature achieved after equilibration is T~1420 K.

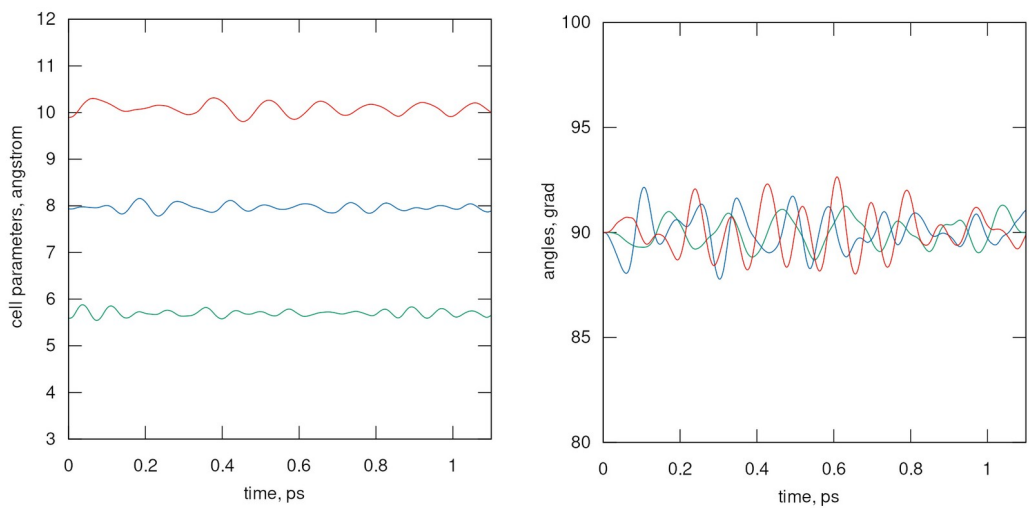

Figure 10. Behavior of crystal cell parameters for $ScC_2B_{12}$ compound obtained in Born-Oppenheimer molecular dynamics simulations.

The conventional method to check the dynamical stability is based on choosing the normal phonon mode representation for the description of dynamics. The unstable modes can be identified then by the presence of imaginary frequencies in phonon dispersion curves (see, for instance, [22, 23]). This approach has been used in our previous paper to establish the dynamical stability of Ca- and Sr- based phases [11].

At the same time, the dynamical stability of a system established by examining the behavior of normal phonon modes is based on harmonic approximation. Alternatively, more general and even more accurate approach would be to use ion molecular dynamics (MD) to directly examine thermalization of a system at temperatures close or above the Debye temperature.

We employed this approach to verify the thermal stability of Sc- and Y- based phases under consideration. Computer simulations were performed on the basis of Born-Oppenheimer molecular dynamics within the framework of general DFT approach used in this work. The simulations were performed for zero pressure for free cell geometry, i.e., the dynamics of cell dimensions and angles was taken into account as well. The initial temperatures were set close to Debye temperatures for the respective compounds; the total time span was ~1.1 -1.5 ps; the time step was set 0.0005 ps. As an example, we give the results for the phase $ScC_2B_{12}$ in Figs. 9-10.

The results of simulations indicate that, after thermalization period of ~0.4-0.5 ps, the system remains in steady state. The equilibration time depends on the initial system state. Upon thermalization, the system retains its initial orthorhombic crystal system and structure, i.e., no any phase change occurs. This is evident also from the direct visual observations of configurations in the process of MD simulations. Therefore, one should conclude that the phase $ScC_2B_{12}$ is thermally (and, therefore, dynamically) stable. The MD simulations preformed for the phase $YC_2B_{12}$ yielded similar results demonstrating its thermal stability.

## IV. Conclusions

First principle DFT simulations are employed to study structural and mechanical properties of orthorhombic $B_{12}$-based metal borides and carbo-borides. The simulations predict the existence of a new family of superhard $AlMgB_{14}$ -related orthorhombic phases ($MeC_2B_{12}$; Me= Mg, Ca, Sr, Sc, Y) with space group Imma [74] and similar structure. Thermal (dynamical) stability of these phases is demonstrated. The calculated isotropic shear and Young's moduli of predicted phases are within the range 230-250 and 530-550 GPa, respectively, and estimated Vickers hardness is 35-55 GPa. These results are supported by direct simulations of shear strength in different directions, which indicate up to the 30-50% increase as compared to the reference $AlMgB_{14}$ compound. Due to distinguishing mechanical properties, this theoretically predicted family of compounds may be promising for creating novel superhard materials.

## Acknowledgment

The authors gratefully acknowledge the support of this research within the framework of the CAS President's International Fellowship Initiative, grant No 2020VEB0005.